\documentclass[pra,twocolumn,
superscriptaddress,showpacs]{revtex4}
\usepackage{amsmath}
\usepackage{amsfonts}
\usepackage{graphicx}

\errorstopmode

\begin{document}

\title{How to share a continuous-variable quantum secret by optical
interferometry}

\author{Tom\'a\v{s} Tyc \cite{Tyc} and Barry C.\ Sanders}

\affiliation{Department of Physics, Macquarie University, 
Sydney, New South Wales 2109, Australia}

\date{November 21, 2001}

\begin{abstract}

We develop the theory of continuous-variable quantum secret sharing and
propose its interferometric realization using passive and active
optical elements.  In the ideal case of infinite squeezing, a fidelity
${\cal F}$ of unity can be achieved with respect to reconstructing the
quantum secret. We quantify the reduction in fidelity for the (2,3)
threshold scheme due to finite squeezing and establish the condition 
for verifying that genuine quantum secret sharing has occurred.

\end{abstract}
\maketitle

\section{Introduction}

Classical secret sharing (CSS) introduced by Shamir \cite{Sha79}
is an important primitive of protection of a secret classical information.
It involves a dealer who distributes a secret amongst a group of $n$ parties
(players)  in a way that prevents all unauthorized subsets of players
(referred to as the adversary structure) from reconstructing this state
and permits authorized subsets (referred to as the access structure) to
successfully reconstruct the state.

In quantum information theory, where the quantum state itself is a
repository of information, protection of this state is of
paramount importance. The quantum version of secret sharing (QSS)
\cite{Cle99,Got00,Smi00} will likely
play a key role in protecting secret quantum information,
e.g.~in secure operations of distributed quantum computation,
sharing difficult--to--construct ancilla states and joint sharing
of quantum money, as examples of the versatility of QSS~\cite{Got00}.
Other examples of quantum security include quantum key
distribution \cite{Ben84}, protection of a classical secret
by quantum means in the presence of eavesdroppers \cite{Hil99,Tit01},
and quantum bit commitment \cite{QBC}.

Whereas quantum secret sharing has been developed for discrete variables
\cite{Cle99,Got00,Smi00}, here we develop continuous-variable (CV) quantum
secret sharing and 
show how it can be implemented using optical interferometry and squeezed light
sources~\cite{Wal86}. Both the theory and the proposed experimental
realization are quite different from their discrete variable counterparts,
yet serve the name goal: sharing secret quantum states.
CV quantum information theory has achieved enormous
success in quantum teleportation~\cite{Vai94} and quantum
computation~\cite{Llo99}, and continuous-variable quantum secret sharing
can be expected to play a key role in future integrated CV quantum
information systems. The recent explosion of work
going into linear optical quantum computation is an example of the
importance of interferometric approaches to quantum information \cite{Kni01},
and our proposal fits well with this rapidly growing
subfield of quantum information research.

We develop the $(2,2)$ threshold quantum secret sharing protocol for
passive linear interferometry, and our protocols for higher threshold QSS
schemes require squeezed light.  We show that finite squeezing limits
the fidelity of the reconstructed secret quantum state, and we
establish a lower bound on fidelity for the $(2,3)$ threshold scheme.
This lower bound is a criterion for establishing that genuine QSS
(as opposed to classical secret sharing) has been achieved.
Finally we discuss the extension of our protocol to non-threshold QSS
schemes.

In contrast to the case of discrete-variable
QSS~\cite{Cle99,Got00,Smi00}, continuous-variable QSS involves
states in infinite dimensional Hilbert spaces.  Quantum
information theory in infinite dimensional Hilbert spaces is a feature
of working with continuous variables~\cite{Vai94,Llo99}.
In the QSS scheme discussed here, the secret state
$\psi$ from an infinite-dimensional Hilbert space $\mathcal{H}$
is encoded into an entangled state $\Phi\in\mathcal{H}^{\otimes n}$
as $n$ shares, one for each player. 
The entanglement is designed so that entanglement swapping operations
\cite{Zuk93} by authorized groups of players can recover the secret state, and
unauthorized groups recover no information whatsoever about the
secret state.

\section{Threshold schemes}

We will consider first the threshold quantum secret sharing scheme for
CV.  Whereas secret sharing is concerned with general
adversary and access structures, threshold secret sharing considers a
particular access structure.  For~$n$ players, the access structure
for $(k,n)$ threshold secret sharing~\cite{Cle99} is the
set consisting of all groupings of~$k$ or more players, and the
adversary structure is the set of all groupings consisting of
fewer than~$k$ players.  We will discuss in detail
the $(k,2k-1)$ threshold scheme. The
general $(k,n)$ scheme, with $n \leq 2k-1$, can be
achieved from $(k,2k-1)$ scheme by having the dealer
discard $2k-n-1$ shares prior to dealing the
state~$\Phi$, and threshold schemes with $n\geq 2k$ are not possible due to
the non-cloning theorem \cite{Cle99,Woo82}.
In the $(k,2k-1)$ threshold scheme, the
dealer's state~$\Phi$ may be operated on by $k$ collaborators to produce
the output state $\Phi_{\rm out}\in\mathcal{H}^{\otimes 2k-1}$.
Ideally, $\Phi_{\rm out}$
is a product of the original secret state~$\psi$ and $k-1$ pairwise
entangled states, each in the Hilbert space
$\mathcal{H}_i\otimes\mathcal{H}_j$,
where the $i^{\rm th}$ and $j^{\rm th}$ player is a collaborator and adversary
(or non-collaborating player), respectively. This entanglement of states
between collaborator and adversary ensures that adversaries do not acquire any
information whatsoever about~$\psi$.

We introduce a continuous-variable representation for the secret state
$\vert\psi\rangle$ as $\psi(x)=\langle x \vert \psi \rangle$ for~$\vert
x \rangle$ the eigenstate of the canonical position operator~$\hat{x}$,
and the eigenvalue spectrum is $x \in \mathbb{R}$.  The
eigenstates of~$\hat{x}$ are not normalizable but satisfy
the orthogonality relation
\begin{equation}
 \langle x^\prime \vert x \rangle = \delta(x-x^\prime).
\end{equation}
In an optical system,
this is the quadrature-phase representation, which can be measured via
optical homodyne detection~\cite{Yue78}.

For ${\bf x}\equiv(x_1,\ldots,x_k)^{\rm T}$ a vector from the $k$-dimensional
vector space $\mathbb{R}^k$ for the canonical positions of $k$ players,
the dealer implements a particular linear mapping
\begin{equation}
\label{mapping:L}
L:\mathbb{R}^k \rightarrow \mathbb{R}^{2k}: {\bf x}
	\mapsto L({\bf x})=\left( x_1,L_1({\bf x}),
	\ldots ,L_{2k-1}({\bf x})\right)^{\rm T}~.
\end{equation}
The linear mapping~$L$ is constrained by the requirement (which can always
be satisfied~\cite{Cle99}) that the components of any $k$-element subset
of $\{x_1,L_1,\ldots,L_{2k-1}\}$ are linearly independent.
The mapping~$L$ is used by the dealer to encode~$\psi$ into the entangled state
\begin{equation}
\vert\Phi\rangle
	= \int_{\mathbb{R}^k} \psi(x_1)\,
   \vert L_1({\bf x})\rangle_1  \cdots \vert L_{2k-1}({\bf x})\rangle_{2k-1}
       {\rm d}^k {\bf x}~
\label{encoding}
\end{equation}
(which is not normalizable for the same reason that~$\vert x \rangle$
is not normalizable).

The encoding~(\ref{encoding}) enables $\psi$ to be reconstructed from any
$k$ shares as follows. Let~$(r_1,\ldots,r_{2k-1})$ be an arbitrary
permutation of indices $(1,2,\ldots,2k-1)$. As both sets
$\{L_{r_1},L_{r_2},\dots,L_{r_k}\}$
and $\{x_1,L_1,\ldots,L_{2k-1}\}$ are linearly independent,
there exists a non-singular $k \times k$ matrix~$T$ such that
\begin{equation}
\label{mapping:T}
T\left(\begin{array}{c} L_{r_1}\\ L_{r_2}\\ \vdots \\ L_{r_k}
    \end{array}\right)
   =\left(\begin{array}{c} x_1\\ L_{r_{k+1}}\\ \vdots \\ L_{r_{2k-1}}
    \end{array}\right)~.
\end{equation}
Given~$T$, there exists a unitary operator $U(T)$ such that
\begin{multline}
 U(T)\vert L_{r_{1}}
 \rangle_{r_1}\vert L_{r_{2}}\rangle_{r_2}\cdots|L_{r_{k}}\rangle_{r_k} \\
=\vert\vert T \vert\vert^{1/2}\,|x_1\rangle_{r_1}\vert L_{r_{k+1}}\rangle_{r_2}
  \cdots\vert L_{r_{2k-1}}\rangle_{r_k}
\label{unitrans}\end{multline}
with $\vert\vert T \vert\vert=\vert\det T\vert$.
The matrix elements of $U$ in the
continuous basis
\begin{equation}
\{\vert {\bf x}^\prime\rangle\equiv
\vert x^\prime_1 \rangle_{r_1}\cdots|x_k'\rangle_{r_k}\}
\end{equation}
are 
\begin{equation}
\langle {\bf x}^\prime|U|{\bf x}^{\prime\prime}\rangle
     =\vert\vert T \vert\vert^{1/2}\,\prod_{i=1}^k
	 \delta\biggl(\sum_{j=1}^k T_{ij}x''_j-x'_i\biggr),
\label{Uelem}
\end{equation}
with $\{T_{ij}\}$ the matrix elements of~$T$.

\begin{figure}
\includegraphics*[width=3in,keepaspectratio]{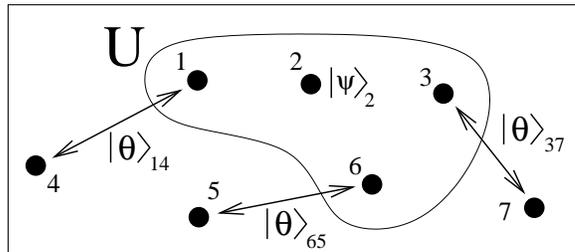}
\caption{The reconstruction of the secret for $k=4$ corresponding to the
permutation $(r_i)=(2136475)$. In order to reconstruct the secret
$\psi$, the players 2, 1, 3 and 6 perform the unitary operation
$U$ from Eq.~(\ref{unitrans}) on their shares. This results in the second
share left in the state $\psi$, while the shares 1, 3, and 6
form EPR states with the shares 4, 7 and 5, respectively. }
\label{reconst}
\end{figure}

The collaborators with shares indexed by $r_1,r_2,\dots,r_k$
reconstruct the secret by transforming their shares via~$U$,
which results in the total state of all shares
\begin{multline}
U\vert\Phi\rangle
  =J \vert\vert T \vert\vert^{1/2}
  \int_{\mathbb{R}^k} \psi(x_1)\,|x_1\rangle_{r_1}  \\ \times
  \vert L_{r_{k+1}}\rangle_{r_2}\cdots
   \vert L_{r_{2k-1}}\rangle_{r_k} 
   \vert L_{r_{k+1}}\rangle_{r_{k+1}}\cdots
   \vert L_{r_{2k-1}}\rangle_{r_{2k-1}}  \\ \times
       {\rm d} x_1{\rm d} L_{r_{k+1}}\cdots{\rm d} L_{r_{2k-1}}  \\
  =J \vert\vert T \vert\vert^{1/2}\vert\psi\rangle_{r_1} \,
  \vert\Theta\rangle_{r_2,r_{k+1}} \vert\Theta\rangle_{r_3,r_{k+2}}
  \cdots \vert\Theta\rangle_{r_k,r_{2k-1}}
\label{product}
\end{multline}
with~$J$ the Jacobian for the transformation from~$\bf x$
to $(x_1,L_{r_{k+1}},\dots,L_{r_{2k-1}})$ and
\begin{equation}
 \vert\Theta\rangle_{ij}\equiv
 \int_{\mathbb R}\vert x\rangle_i\vert x\rangle_j\,{\rm d} x.
\label{Theta}
\end{equation}
Equation~(\ref{product}) shows that the $r_1{}^{\rm th}$ share is the secret
state $\psi$ and shares $r_2,\dots,r_k$ 
are maximally entangled with the shares of the adversaries (see
Fig.~\ref{reconst}). Thus the quantum secret is reconstructed from any $k$
shares via a unitary transformation, and any $k-1$ shares produces no
information about~$\psi$ whatsoever as tracing over the remaining $k$ shares
yields a multiple of the identity operator.

Important components of the state~(\ref{product}) are the (unnormalized and
ideal) EPR states~\cite{Ein33} $\vert\Theta\rangle_{ij}$
(see Eq.~(\ref{Theta})) such that
$_{ij}\langle x \, x^\prime \vert \Theta \rangle_{ij} = \delta(x-x^\prime)$.
These states can be approximated by the strongly squeezed two-mode vacuum
states~\cite{Sch85} $\vert\eta\rangle_{ij}$ that have the representation
\begin{equation}
{}_{ij}\langle x \, x^\prime \vert\eta\rangle_{ij}
	\equiv(1-\eta^2)^{1/2}\sum_{n=0}^\infty
		\eta^n u_n(x) u_n(x^\prime)
\end{equation}
for~$-1\le\eta\le 1$ and $u_n(x)\equiv\langle x\vert n\rangle$ with
$\vert n\rangle$ the Fock state.
As
\begin{equation}
 \sum_{n=0}^\infty u_n(x)u_n(x^\prime)=\delta(x-x^\prime),
\end{equation}
it holds
\begin{equation}
 \lim_{\eta\rightarrow 1}(1-\eta^2)^{-1/2}\langle xx^\prime\vert\eta\rangle
 =\delta(x-x^\prime)
\end{equation} and hence
$\vert\eta\rightarrow1\rangle=\vert\Theta\rangle$. 
The share $i$ of $\vert\eta\rangle_{ij}$, after tracing over the
share $j$, behaves locally as a thermal state $\varrho_T$ with temperature
$T=-\hbar\omega/(2k_B\ln|\eta|)$, where $k_B$ denotes the
Boltzmann constant; the limit $\eta\rightarrow 1$ corresponds to
$T\rightarrow \infty$.

\section{Experimental realization}

Now we show how to implement the above reconstruction procedure experimentally.
Given $T$ in Eq.~(\ref{mapping:T}), there exists $S\in{\rm Sp}(2k,\mathbb{R})$
corresponding to $T$ mapping the canonical position and $(T^{\rm T})^{-1}$
mapping the corresponding canonical momenta, for $^{\rm T}$ the transpose.
The symplectic transformation $S$ can be decomposed into a sequence of
SU(2) and SU(1,1) transformations \cite{Bar01} that perform passive
and active (squeezing) operations on the shares.
Physically, such transformations can be realized by the combined
SU(2) and SU(1,1) interferometry \cite{Yur86} that uses
beam splitters, mirrors, phase shifters and squeezers. The secret and shares
are then distinct spatial modes of light, and
an interferometer can be designed by any group of $k$ collaborators to
yield the secret at one output port when the shares are injected to the
input ports.

The encoding of the secret by the dealer can also be performed via (active)
interferometry (again a symplectic
transformation). For a chosen permutation of indices, the dealer
creates the state (\ref{product}) and employing a suitable interferometer
applies $U^\dagger$ to this state in order 
to obtain $\Phi$. In summary, the dealer can encode the secret state $\psi$
in a $2k-1$ mode entangled state $\Phi$ via interferometry which can be
decoded by any $k$ collaborators also by interferometry. As has been
mentioned at the start, the general $(k,n)$ threshold scheme can be achieved
by having the dealer discard $2k-1-n$ shares.

\section{Example: (2,3) threshold scheme}

\begin{figure}
\includegraphics*[width=3in,keepaspectratio]{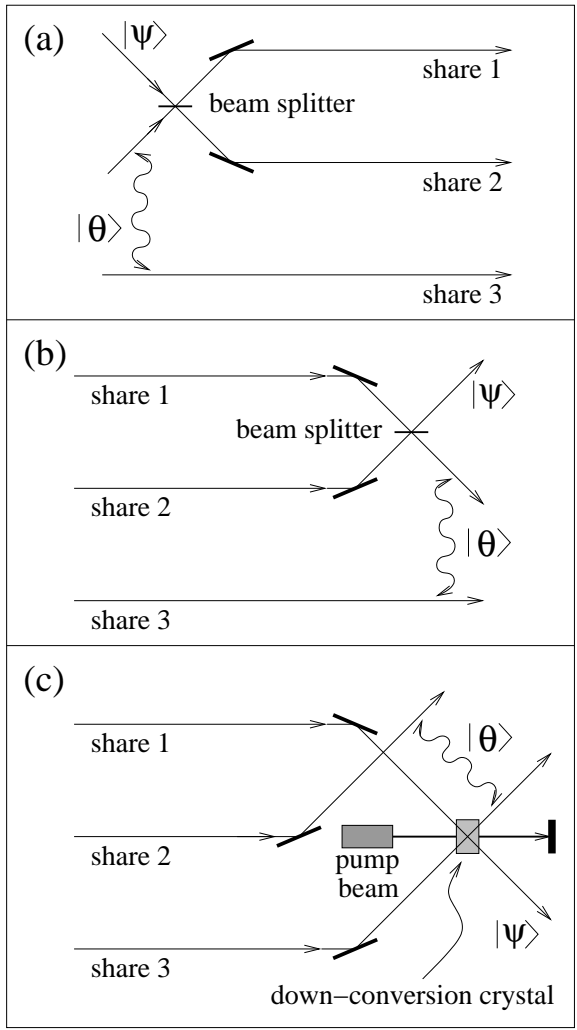}
\caption{Encoding and reconstruction procedures for a $(2,3)$ threshold scheme.
(a)~For the encoding process, the dealer creates the state
$|\psi\rangle_1|\Theta\rangle_{23}$ and combines modes 1 and 2 on a 50/50 BS;
(b)~Players 1 and 2 combine their shares on a 50/50 BS
    to obtain $\psi$ at one output;
(c)~players 1 and 3 combine their shares at a non-degenerate parametric
down-converter which is pumped by a coherent beam of doubled frequency
to obtain $\psi$ at one output.
In both cases (a) and (b) the remaining output forms an EPR state with the
share of the adversary.}
\label{23scheme}
\end{figure}

We will give an example of the $(2,3)$ threshold scheme. The dealer chooses
\begin{equation}
L_1=\frac{x_2+x_1}{\sqrt2}, \quad L_2=\frac{x_2-x_1}{\sqrt2}, \quad L_3=x_2,
\end{equation}
and constructs the
corresponding interferometer with a 50/50 beam splitter (BS) as shown in
Fig.~\ref{23scheme}~(a). This interferometer transforms the initial state
$|\psi\rangle_1|\Theta\rangle_{23}$ to the three-mode entangled state
\begin{equation}
\vert\Phi\rangle=
\int_{{\mathbb R}^2} \psi(x_1)\,\left\vert\frac{x_2+x_1}{\sqrt2}\right\rangle_1
         \left\vert\frac{x_2-x_1}{\sqrt2}\right\rangle_2
         \left\vert x_2\right\rangle_3 \,{\rm d} x_1\,{\rm d} x_2 .
\label{encoding2}
\end{equation}
The secret can then be reconstructed from any two shares.
By combining shares 1 and 2 on a 50/50 BS (Fig.~\ref{23scheme}~(b)),
thereby transforming the canonical positions of the two shares via 
\begin{equation}
T_{12}=\frac1{\sqrt2}\left(\begin{array}{rr} 1 & -1\\ 1 & 1 \end{array}\right)
\end{equation}
the first share is left in the secret state $\psi$.
Similarly by combining the first and third shares on a non-degenerate
parametric down-conversion crystal, pumped by a coherent field
(see Fig.~\ref{23scheme}~(c)), that transforms the canonical positions via 
\begin{equation}
T_{13}=\left(\begin{array}{rr}\sqrt2 & -1\\ -1 & \sqrt2 \end{array}\right),
\end{equation}
the first share is left in the secret state
$\psi$. A similar procedure can be employed for shares 2 and 3 to
reconstruct the secret.

\begin{figure}
\includegraphics*[width=3in,keepaspectratio]{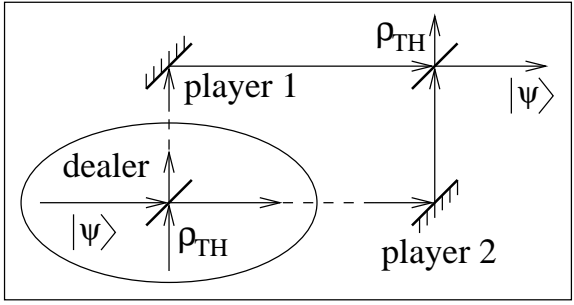}
\caption{The $(2,2)$ threshold scheme: the dealer encodes the secret
by combining it with a thermal state of an infinite temprerature on a 50/50 BS.
The two players can then reconstruct the secret by combining their
beams on another 50/50 BS.}
\label{22scheme}
\end{figure}

A surprisingly simple $(2,2)$ quantum secret sharing threshold scheme can be
derived from the previous $(2,3)$ scheme by discarding the third share.
The (unnormalized) reduced density operator of the first two shares after
tracing over the last share is
\begin{equation}
\rho_{12}= U^\dagger(|\psi\rangle_1\langle\psi|\otimes
   \openone_2)U,
\end{equation}
for $U$ the unitary operator induced by $T_{12}$ according to
Eqs.~(\ref{unitrans}) and (\ref{Uelem}), and $\openone$ the unit operator.
As the thermal state $\varrho_{T\rightarrow\infty}$
is a multiple of the unit operator, the state of the two shares can be
also obtained by mixing the secret $\psi$ with a thermal state of infinite
temperature on a 50/50 BS (see Fig.~\ref{22scheme}).
The encoding of the secret is thus performed by simply mixing it
with a thermal state on a BS, whereas the reconstruction
is accomplished via recombining the two shares on another 50/50 BS.
As a result, the players obtain the secret state as well as the thermal state
from the two BS output ports.

\section{The case of finite squeezing}

\begin{figure}
\includegraphics*[width=3in,keepaspectratio]{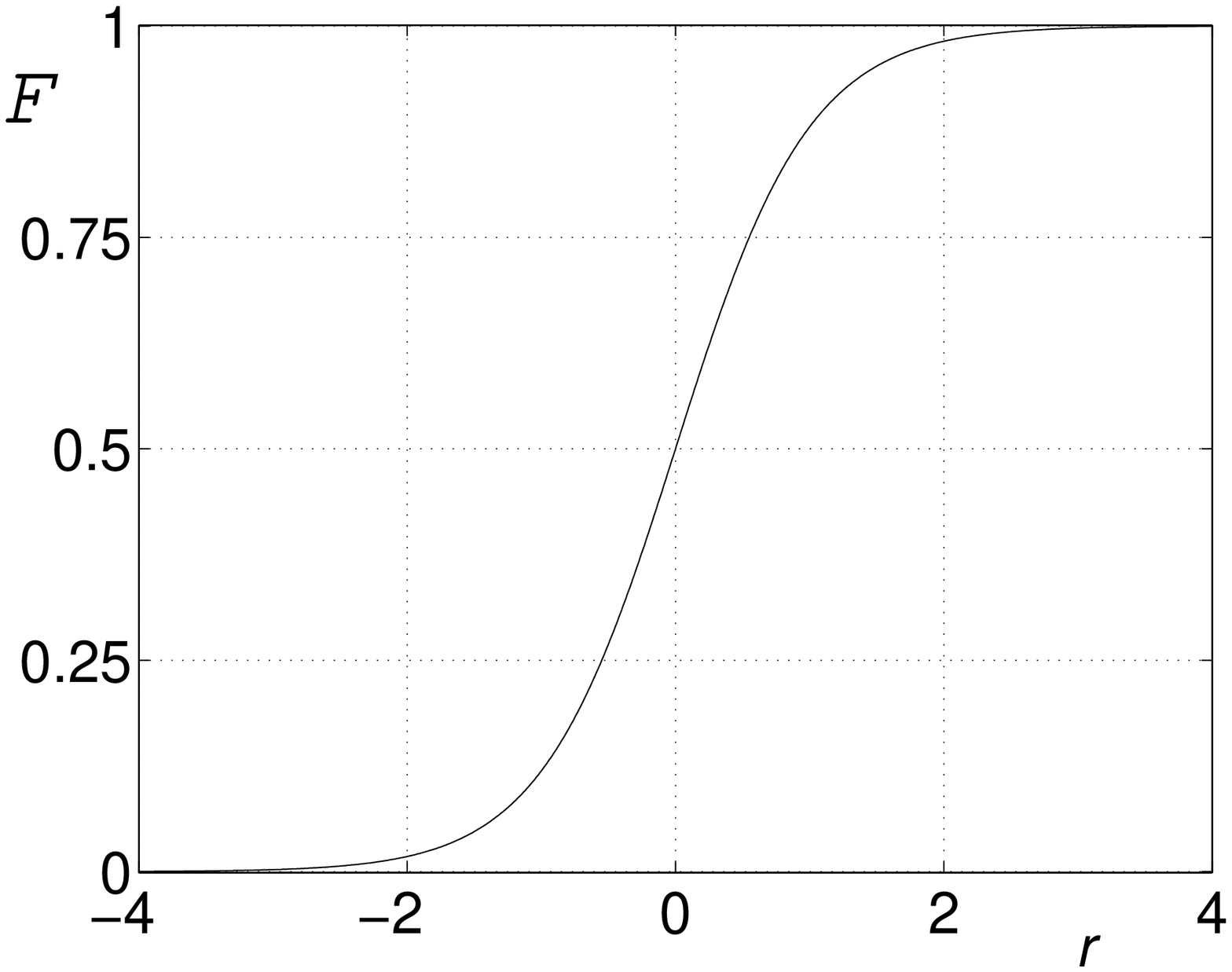}
\caption{The fidelity ${\cal F}=\langle\psi\vert\rho|\psi\rangle$
for the scheme in Fig.~\ref{23scheme}~(a),~(c) with squeezing parameter
$r$ used by the dealer and the secret $\psi$ a coherent state.}
\label{fidelity}
\end{figure}

Whereas infinite temperature is necessary for ideal $(2,2)$
quantum secret sharing threshold scheme, $\varrho_{T\rightarrow\infty}$
contains on average an infinite number of photons so any finite error
in the encoding or reconstruction process will produce
infinitely many photons from output port~1 of the BS
and thereby destroy the secret completely.
Therefore a finite-temperature $\varrho_T$ must be employed instead, and
in the general case of $(k,n)$ threshold scheme, finitely squeezed two-mode
vacuum states must replace EPR states for the same reason.
This will generally compromise the secret sharing fidelity, i.e., the
overlap of the reconstructed state and the original secret state
\begin{equation}
 {\cal F}=\langle\psi|\rho|\psi\rangle,
\end{equation}
where $\rho$ is the (generally mixed) state obtained as the result of
reconstruction. Consider a dealer using an interferometer as in
Fig.~\ref{23scheme}~(a) for the encoding process, where a two-mode squeezed
vacuum state $|\eta=\tanh r\rangle$ replaces the EPR state at the two
inputs. Clearly, players 1 and 2 can still reconstruct the secret perfectly
while players 1 and 3 or players 2 and 3 cannot.
If the secret is a coherent state, the fidelity for players 1 and 3 is
\begin{equation}
 {\cal F}=\frac1{1+{\rm e}^{-2r}}=\frac{1+\eta}2
\end{equation}
(see Fig.~\ref{fidelity}). If $r=0$ (no squeezing),
the players can still achieve ${\cal F}=1/2$.
This fidelity threshold can be used to verify whether a genuine quantum secret
sharing has taken place in this particular $(2,3)$ threshold scheme.

\section{Conclusion}

To conclude, we have developed a theory of quantum secret sharing using
CV and shown how encoding and reconstruction processes
could be achieved via (active) multimode interferometry for the $(k,n)$
threshold scheme. The $(2,3)$ and $(2,2)$ schemes
have been presented in detail, including an allowance for finite
squeezing and a minimum fidelity necessary to demonstrate that genuine
secret sharing has been performed.
The $(2,2)$ threshold scheme is achievable with current technology.

The $(k,n)$ threshold scheme is readily generalized to an arbitrary adversary
structure $\cal A$ by analogy with the discrete-variable schemes 
based on monotone span programs \cite{Smi00}.
For any adversary structure $\cal A$, there exists a self-dual structure
$\cal A'$ ($\cal A'$ is self-dual iff, for any division of the set of all
players into two disjoint groups, exactly one group is able to reconstruct
the secret), from which $\cal A$ can be obtained by discarding some
shares \cite{Smi00}.
For the self-dual adversary structure $\cal A'$,
the encoding procedure of the dealer and decoding procedures of the
collaborating players can again be realized by linear
mappings of the canonical positions by employing a suitable interferometer.
For a total of $n$ players, the initial state of the dealer consists of the
secret $\psi$ and $n-1$ single-mode infinitely squeezed vacuum states
\cite{Wal86}. 

\acknowledgments
This project has been supported by a Macquarie University Research
Grant and by an Australian Research Council Large Grant.  We appreciate
valuable discussions with J. Pieprzyk and I.\ Shparlinski.

\newpage

\begin{thebibliography}{99}
\bibitem[*]{Tyc} On leave from Institute of Theoretical
Physics, Masaryk University, Kotl\'a\v rsk\'a 2, 61137 Brno, Czech Republic;
Email: tomtyc@physics.muni.cz
\bibitem {Sha79} A.\ Shamir, Comm.\ of the ACM \textbf{22}, 612 (1979).
\bibitem {Cle99} R.~Cleve et al, \prl \textbf{83}, 648 (1999).
\bibitem {Got00} D.~Gottesman, \pra \textbf{61}, 042311 (2000).
\bibitem {Smi00} A.\ D.\ Smith, quant-ph/0001087.
\bibitem {Ben84} C.~H.~Bennett and G.~Brassard, in {\em Proc. of IEEE
Inter. Conf. on Computers, Systems and Signal Processing,
Bangalore, India} (IEEE, New York, 1984), p.~175.
\bibitem {Hil99} M.\ Hillery et al, \pra \textbf{59}, 1829 (1999).
\bibitem {Tit01} W.\ Tittel et al, \pra \textbf{63}, 042301 (2001).
\bibitem {QBC} Although quantum bit commitment cannot be unconditionally
               secure [H.-K.\ Lo and H.\ F.\ Chau,
	\prl \textbf{78}, 3410 (1997); D. Mayers, \prl \textbf{78}, 3414 (1997)], nontrivial bounds on
	cheating by either sender or receiver have been established [R.\ W.\ Spekkens and T.\ Rudolph, 
	quant-ph/0106019 and quant-ph/0107042].  These bounds are significant, as bit commitment,
	along with secret sharing, is an important primitive in data security.
\bibitem {Wal86} D.\ F.\ Walls, Nature \textbf{324}, 210 (1986).
\bibitem {Vai94} L.\ Vaidman, \pra \textbf{49}, 1473 (1994);
             S.~L.~Braunstein and H.~J.~Kimble, \prl \textbf{80}, 869 (1998);
                 A.\ Furusawa et al, Science {\bf 282}, 706 (1998).
\bibitem {Llo99} S.\ Lloyd and S.\ L.\ Braunstein, \prl {\bf 82}, 1784 (1999).
\bibitem {Kni01} E.\ Knill et al, Nature \textbf{409}, 46 (2001).
\bibitem {Zuk93} M.\ \.{Z}ukowski et al, \prl \textbf{71}, 4287 (1993).
\bibitem {Woo82} W.~K.~Wootters and W.~H.~Zurek, Nature {\bf 299}, 802 (1982).
\bibitem {Yue78} H.~P.~Yuen and J.~H.~Shapiro, IEEE~Trans.~Inf.~Theory
                 \textbf{IT-25}, 179 (1979); {\bf IT-26}, 78 (1980).
\bibitem {Ein33} A.\ Einstein et al, Phys.\ Rev.\ {\bf 47},
                 777 (1935).
\bibitem {Sch85} B.\ L.\ Schumaker and C.\ M.\ Caves,
                 \pra {\bf 31}, 3093 (1985).
\bibitem {Bar01} S.~D.~Bartlett et al, \pra \textbf{63}, 42310 (2001).
\bibitem {Yur86} B.~Yurke et al, \pra \textbf{33}, 4033 (1986).





\end{thebibliography}
\end{document}